\pdfoutput=1

\documentclass[prd,amsmath,amssymb,nofootinbib,preprintnumbers]{revtex4}

\usepackage{color}
\usepackage{float}
\usepackage{pdfpages}
\usepackage{gensymb}

\def\apjs{ApJS}
\def\aap{A\&A}
\def\jcap{J. Cosmology Astropart. Phys.}
\def\prd{Phys.~Rev.~D}

\def\lsim{\mathrel{\raise.3ex\hbox{$<$\kern-.75em\lower1ex\hbox{$\sim$}}}}
\def\gsim{\mathrel{\raise.3ex\hbox{$>$\kern-.75em\lower1ex\hbox{$\sim$}}}}

\def\beq{\begin{equation}}
\def\eeq{\end{equation}}
\def\ave{\text{ave}}
\def \newlinefix {\hfill\break}
\def \GCDMCITE   {\cite{GCDMBounds1} \cite{GCDMBounds2}}

\newcommand{\like}{\ensuremath{\mathcal{L}}}
\newcommand{\mchi}{\ensuremath{m_{\chi}}}
\newcommand{\sigmav}{\ensuremath{\langle\sigma v\rangle}}
\newcommand{\hatsigmav}{\ensuremath{\widehat{\langle\sigma v\rangle}}}

\newcommand{\dPhippdE}{\ensuremath{\frac{d\Phi_{\makebox[0pt][l]{\tiny pp}}}{dE}\phantom{i}}}

\newcommand{\hatJX}{\ensuremath{\widehat{(\mathcal{JX})}}}
\newcommand{\hatbeta}{\ensuremath{\widehat{\mathbf{\beta}}}}
\newcommand{\hatgamma}{\ensuremath{\widehat{\mathbf{\gamma}}}}

\newcommand{\dhatJX}{\ensuremath{\widehat{\vphantom{\rule{0pt}{2.3ex}}\smash{\widehat{(\mathcal{JX})}}}}}
\newcommand{\dhatbeta}{\ensuremath{\widehat{\vphantom{\rule{-1ex}{2.1ex}}\smash{\widehat{\mathbf{\beta}}}}}}
\newcommand{\dhatgamma}{\ensuremath{\widehat{\vphantom{\rule{0pt}{1.5ex}}\smash{\widehat{\mathbf{\gamma}}}}}}

\def\beq{\begin{equation}}
\def\eeq{\end{equation}}
\def\bey{\begin{eqnarray}}
\def\eey{\end{eqnarray}}
\def\textbfig{\begin{figure}}
\def\efig{\end{figure}}

\def\lsim{\mathrel{\raise.3ex\hbox{$<$\kern-.75em\lower1ex\hbox{$\sim$}}}}
\def\gsim{\mathrel{\raise.3ex\hbox{$>$\kern-.75em\lower1ex\hbox{$\sim$}}}}

\def\GeV{\, {\rm GeV}}

\def\lsim{\mathrel{\raise.3ex\hbox{$<$\kern-.75em\lower1ex\hbox{$\sim$}}}}
\def\gsim{\mathrel{\raise.3ex\hbox{$>$\kern-.75em\lower1ex\hbox{$\sim$}}}}

\def\beq{\begin{equation}}
\def\eeq{\end{equation}}
\def\ave{\text{ave}}
\def \newlinefix {\hfill\break}

\begin{document}

\title{ Improved Constraints on Dark Matter Annihilation to a Line \protect\\
    using Fermi-LAT observations of Galaxy Clusters  }

\author{Douglas Quincy Adams}
\email{doug.q.adams@gmail.com}
\affiliation{
Department of Astrophysical Sciences,
Princeton University,
Payton Hall,
1 Ivy Ln, Princeton, NJ 08540, USA
}

\author{Lars Bergstrom}
\email{lbe@fysik.su.se}
\affiliation{
Oskar Klein Centre for Cosmoparticle Physics, 
Department of Physics,
Stockholm University, 
AlbaNova University Center,
SE-106 91 Stockholm, Sweden}

\author{Douglas Spolyar}
\email{dspolyar@gmail.com}
\affiliation{
Oskar Klein Centre for Cosmoparticle Physics, 
Department of Physics,\\
Nordita (Nordic Institute for Theoretical Physics)\\
Stockholm University, 
AlbaNova University Center,
SE-106 91 Stockholm, Sweden
}
%

\date{\today}


\begin{abstract}

    Galaxy clusters are dominated by dark matter, and may have a larger proportion of surviving substructure than, e.g, field galaxies.
    Due to the presence of galaxy clusters in relative proximity and their high dark matter content, 
    they are promising targets for the indirect detection of dark matter via $\gamma$ rays. 
    Indeed, dedicated studies of sets of up to 100 clusters have been made previously, so far with no clear indication of a dark matter signal. 
    Here we report on $\gamma$-ray observations of some 26,000 galaxy clusters based on Pass-7 Fermi Large Area Telescope (LAT) data, 
    with clusters selected from the Tully 2MASS Groups catalog. 
    None of these clusters is significantly detected in $\gamma$ rays, and we present $\gamma$-ray flux upper limits between 20 GeV and 500 GeV. 
    
    We estimate the dark matter content of each of the clusters in these catalogs, and constrain the dark matter annihilation cross section, 
    by analyzing Fermi-LAT data from the directions of the clusters. 
    We set some of the tightest cluster-based constraints to date on the annihilation of dark matter particles with masses between 20 GeV and 500 GeV for annihilation to a gamma-ray line. 
    Our cluster based constraints are not yet as strong as bounds placed using the Galactic Center, 
    although an uncertainty still exists regarding the ``boost factor'' from cluster substructure, where we have chosen a rather conservative value.
    Our analysis, given this choice of possible boost, is not yet sensitive enough to fully rule out typical realistic DM candidates, especially if the gamma-ray line is not a dominant annihilation mode.

\end{abstract}


\maketitle

\section{Introduction}

Weakly Interacting Massive Particles (WIMPs) (for reviews, see Refs.~\cite{Jungman:1995df,Bergstrom:2000pn,Bertone:2004pz}) 
are thought to be among the best motivated dark matter (DM) candidates.  
Some of these candidates annihilate among themselves in the early universe with a strength to naturally provide the correct
relic density today to explain the dark matter abundance in the universe.  
It is natural to assume then that these DM candidates go through similar annihilation processes in the present universe, 
wherever the DM density is sufficiently high. This is the basis for DM indirect detection experiments, which  
search for the annihilation products of such dark matter particle candidates.
Possible interesting annihilation model final end products include, but are not limited to: electrons and/or positrons, antiprotons, photons, and neutrinos.

Two of the most promising DM annihilation products to observe are neutrinos and photons. 
Neutrinos and photons have the advantage that they are not charged and therefore deflected by magnetic fields in the Galaxy. Therefore directionality for source of origin can be preserved for indirect DM searching.  Recent searches for DM annihilation
to neutrinos have found no excess over background towards the Sun~\cite{2016JCAP...04..022A,2015arXiv150304858T},
 Galactic Center~\cite{2009arXiv0905.4764S,2016arXiv160600209I} , and nearby galaxies and clusters~\cite{2013PhRvD..88l2001A}.
(For problems and possibilities concerning charged antimatter particles, see the reviews~\cite{Jungman:1995df,Bergstrom:2000pn,Bertone:2004pz}). 
Promising sites for photons include 
our Galactic Halo~\cite{Ellis:1988qp,Turner:1989kg,Kamionkowski:1990ty,Silk:1984zy},
the Galactic Center~\cite{Bergstrom:1997fj}, 
dwarf satellite galaxies~\cite{Evans:2003sc,Bergstrom:2005qk}, 
clusters of galaxies \cite{Colafrancesco:2005ji,Pinzke:2009cp,Pinzke:2011ek}, 
and DM substructures~\cite{Bergstrom:1998jj,Bertone:2005xz,Sandick:2010yd,Sandick:2011zs,Lavalle:2012ef}.

The Large Area Telescope of the Fermi Gamma Ray Space Telescope
(Fermi-LAT)~\cite{Collaboration:2009zk} has searched for $\gamma$-rays as a signature of DM annihilation. (The strategy was outlined in \cite{Baltz:2008wd}.) 
The best regions investigated so far using the Fermi Telescope have been those with a large abundance of DM: 
the Galactic Center, clusters, and dwarf galaxies, some of the most dark matter dominated objects known.  
Currently, an excess of $\sim$GeV gamma rays that could be due to DM annihilation has been observed by Fermi-LAT near the Galactic Center (GC)~\cite{0910.2998, 1010.2752, 1012.5839, 1110.0006, 1207.6047, 1302.6589, 1306.5725, 1307.6862, 1312.6671, 1402.4090, 1402.6703, 1406.6948, 1409.0042, 1410.6168, Murgia2014}.   At present the true origin 
of these $\gamma$-rays is uncertain, with alternate explanations including a new population of millisecond pulsars (MSPs)~\cite{1011.4275, 1305.0830, 1309.3428, 1402.4090, 1404.2318, Calore:2014oga, 1407.5625, 1411.2980, 1411.4363,OLeary:2015gfa}, cosmic-ray injection~\cite{Carlson:2014cwa, Petrovic:2014uda}, or unresolved point sources in the Inner Galaxy \cite{Lee:2015fea}.

Other than the Galactic Center, the lack of excess signal  over background is used to place
$\gamma$-ray flux upper limits for energies between 500~MeV and
500~GeV.  These bounds are then used to constrain the dark matter
annihilation for a broad range of particle masses and annihilation
channels.  For current  bounds on DM annihilation, see the Fermi-LAT combined analysis of 
dwarf galaxies \cite{Ackermann:2015zua}, 
VERITAS \cite{Aliu:2012ga}, 
and MAGIC \cite{Aleksic:2013xea} 
observations of Segue~1, as well as results of the H.E.S.S.\ collaboration \cite{Abramowski:2014tra} 
on Sagittarius and other dwarf galaxies.   
 In addition, Ref. \cite{Lopez:2015uma} showed that Fermi/LAT observations of Dwarf Galaxies highly constrain a dark matter interpretation of excess positrons seen in AMS-02\cite{Accardo:2014lma}, HEAT~\cite{Barwick:1997ig}, and 
 PAMELA data~\cite{Adriani:2008zr}.

In this paper, we focus on searches for $\gamma$-ray signatures of DM annihilation from galaxy clusters 
in Fermi-LAT data.  These include the most massive virialized DM structures in the universe and could
produce substantial DM annihilation signals. 
We use a dark matter annihilation model within galaxy clusters from which luminosity is derived from two components. 
The ``smooth" component can be described by a radial model within each cluster.
The ``clumpy" component is described by a large number of subhaloes predicted by numerical simulations to exist within the clusters.
The annihilation signal is likely enhanced by the ``clumpy" component from the substructure, 
and this  is expected to dominate over the smooth component by a large enhancement, or ``boost" factor.
Previous searches finding no evidence for gamma-rays from clusters have been used to set 
bounds on DM annihilation \cite{Aleksic:2009ir, Ackermann:2010rg,Huangetal2012,Han:2012uw}.  
A novel feature in the present analysis is that, whereas previous work examined at most $\sim 100$ 
 clusters in total, we enlarge the sample to encompass tens of thousands of galaxy clusters. 
We also investigate directly these stacked data for structure in the energy distribution from the clusters 
(which all are at low redshift enough to have negligible energy shift compared to the $\sim$10--20\% instrument resolution). We note that our strategy is different from that of recent previous studies, where attention is instead given to angular correlations between galaxy gamma-ray data and data of other wavelengths \cite{cuoco, ando}. It is more similar in spirit to a recent analysis of the Fermi-LAT collaboration where, however, only 16 nearby clusters were studied
\cite{Anderson:2015dpc}. 
 In fact, one of the first analyses of this kind \cite{hektor} claimed observation of a gamma-ray line signal around 130 GeV from galaxy clusters that would have been consistent with the indications from \cite{Bringmann:2012vr,weniger_line}. These interesting indications have, however, not been verified with further Fermi-LAT data \cite{fermi_no_line}.

The motivation 
for our analysis is the large number of galaxy groups  identified by Tully in the 2MASS Redshift 
Survey (2MRS) data \cite{tully}.
We search in the direction of these tens of thousands of galaxy groups (hereafter denoted clusters) for an excess of $\gamma$-rays over expected backgrounds.
In this first paper, we look for a $\gamma$-ray line \cite{Bergstrom:1988fp,Bergstrom:1997fj} where, due to the low velocity of the annihilating particles, the energy of the photons is equal to that of the WIMP mass,
\begin{equation}
    E_\gamma = m_\chi \ \, .
\end{equation}
In a forthcoming paper, we will generalize to searching for gamma-rays from DM annihilation through a variety of other channels in addition to annihilation directly to a line.

For the IR-selected galaxies in the Two Micron All Sky Survey (2MASS), redshifts are provided by the 2MASS Redshift Survey (2MRS)  \cite{2MRS}.  
The 2MRS galaxies have been selected to be at
Galactic latitude  $|b| \geq 5^\circ$ ($\geq 8^\circ$ towards the Bulge).
Previous work studying DM annihilation with  2MASS data includes \cite{Ando:2014aoa}. 
Using a combination of 2-D spatial position from 2MASS together with the redshifts, 
Tully was able to identify galaxies grouped together in clusters \cite{tully}.
Given their redshifts, one knows the distances to the clusters.  
Errors on the distance will be related to redshift space distortions (bulk movement of galaxies) and the uncertainty on the Hubble constant.   
These clusters all have redshifts consistent with being nearby, within 280 Mpc (most within 50-100 Mpc). 
The IR light from galaxies is used to determine their stellar masses, from which one can obtain the total mass using
abundance matching. 
This mass estimation is quite accurate because of the lack of extinction in IR. 
For the largest masses, velocity dispersion can be used to estimate the total mass of the cluster, 
and for smaller groups more detailed kinematics of nearby groups can be used to find empirical estimates \cite{tully}. 
The estimates should be accurate to $\sim 20\%$.  
The clusters in Tully's catalogue have masses in the range $10^{12}-10^{15} M_\odot$.   
With the halo masses and distances provided in \cite{tully}, 
we will thus be able to calculate the predicted $\gamma$-ray flux from dark matter 
(for a given WIMP mass and annihilation cross-section).

Our general approach is as follows.  
For each cluster, we define a region of interest containing 95\% of the expected DM annihilation luminosity.  
We slide a bin of energy range $\sim$ twice the energy resolution of Fermi-LAT 
across our full spectrum of interest,
centered at photon energies from  20 GeV up to 500 GeV in steps of 2 GeV. 
We then add up all the observed photons for all the clusters for each energy window, and compare our observed photon count to the expected count.  
We search for a bump in the observed photon count above an expected power law background, 
i.e. we search for a  line (or internal bremsstrahlung \cite{Bergstrom:1989jr,Bringmann:2012vr}) signal at an energy equal to the WIMP mass due to DM annihilation.  
Since no excess above background is found, we use the null signal to place bounds on the DM annihilation cross 
section as a function of WIMP mass.

\section{Dark Matter Annihilation $\gamma$-ray Flux}

The differential $\gamma$-ray flux $\frac{d\phi_k } { dE d\Omega}$ from DM annihilation in
cluster $k$ can be written as the product of two components, a factor
$\dPhippdE\ $ that encodes all the particle physics and the so-called
J-factor \cite{Bergstrom:1997fj} that contains the astrophysics,

\begin{equation} \label{eqn:dphidE}
  (\mathbf{I}_{m_\chi})_k= \frac{d\phi_k }{dE d\Omega} = \dPhippdE \times J_k
          = \bigg(\frac{1}{4\pi}\frac{\langle \sigma v \rangle}{m_\chi^2}\, \delta(E-m_\chi) \bigg)
            \times\bigg(\int_{\textrm{l.o.s.}} \rho^2(r)\,dr\, \bigg)_k \, .
\end{equation}

Here, $\sigmav$ is the DM annihilation cross section,
$m_\chi$ is the DM mass, 
and the $\delta$ function indicates that we are interested in a line signal.
The J-factor (the term in the second set of parentheses) integrates
the square of the DM density $\rho_\chi$ along the line of sight.
For each cluster we must now obtain an estimate of the total J-factor.
The J-factor has two contributions, the smooth halo component and the substructure (clumpy) component.

\subsubsection{The Smooth Component}

From simulations \cite{Navarro:1996gj,Han:2012uw}, for the smooth component of our DM haloes 
we take the Navarro, Frenk, and White (NFW) density profile 
 \begin{equation}
 \rho_{NFW}(r) = {\rho_s \over { (r/r_s) (1+r/r_s)^2} }
 \end{equation}
 where  the characteristic density $\rho_s$ and radius $r_s$
  are related to halo concentration and virial radius
through the relations, 
$\rho_s=\dfrac{200}{3}\dfrac{c^3\rho_c}{{\rm log}(1+c)-c/(1+c)}$ and
$r_s=r_{200}/c$.  Here $\rho_c$ is the critical density of the universe,
$r_{200}$ is the cluster virial radius within which the average density
is $200\rho_c$, and the concentration parameter $c$ is given by 
\begin{equation}
c=5.74(\frac{M_{200}}{2\times 10^{12}h^{-1}M_\odot})^{-0.097}  
\end{equation}
\cite{Duffy:2008pz}.
Here $M_{200}$ is the virial mass  given by Tully.

The J-factor integrated over a large enough solid angle can be used to define the total flux,
$\mathcal{J}_{int}=\int_{\Delta \Omega} J d\Omega$ (this value would be appropriate for a point source 
approximation).  For the smooth component of a cluster, Han {\em et al.} 
\cite{Han:2012uw} find that (see also \cite{Bergstrom:2005qk})
\begin{equation}
    \begin{aligned}
        \mathcal{J}_{NFW} &= {4 \pi \over 3} {1 \over D_A^2} \rho_s^2 r_s^3 {1 \over 8.5 {\rm kpc}} \biggl({1 \over 0.3 {\rm GeV/cm^3}} \biggr)^2 \
    \end{aligned}
    \label{J_NFW}
\end{equation}
Here $D_A$ is the angular diameter distance of the cluster.

$\mathcal{J}_{NFW}$ is a quantity we will use for cluster selection and normalization for the boost factor described shortly.

\subsubsection{The Clumpy Component}

We must now account for enhancement due to substructure inside the cluster, 
which is expected to account for the majority of the dark matter signal from galaxy clusters \cite{Colafrancesco:2005ji,Pinzke:2009cp,  Pinzke:2011ek,Han:2012uw}. 
While the total sub-halo mass constitutes only 10 to 20 percent of the total halo mass, the DM density inside the subhaloes
is considerably higher density than within the smooth component of the cluster.  
Indeed, in comparing with data, it is the substructure $\mathcal{J}_{sub}$ that is the dominant component in the
last term of Eqn. (\ref{eqn:dphidE}).

We define a boost factor
\begin{equation}
    \begin{aligned}
        B =\mathcal{J}_{sub}/\mathcal{J}_{NFW}
    \end{aligned}
\end{equation}
as the ratio of the total integrated luminosity due to the substructure (subscript $sub$) to the total integrated luminosity due to the smooth NFW component. The notation is as
above, with the total integrated flux from the substructure (over a large enough solid angle) as
\begin{equation}
    \begin{aligned}
        \mathcal{J}_{sub}=\int_{\Delta \Omega} J_{sub} d\Omega
    \end{aligned}
\end{equation}
In the last few years much progress has been made in determining the boost factor of dark matter annihilation in clusters (see \cite{Moline:2016pbm} and references therein). 
The improvements have been a result of improvements in estimating the number of substructures in a halo, 
and the annihilation rate in a give sub-halo(which depends strongly on the sub-halo concentration parameter) \cite{Pinzke:2009cp,Pinzke:2011ek,Han:2012uw}. 
The best-guess boost factor per halo is between about 20 and 50 depending on the mass of the halo in our mass range, with boost increasing with cluster mass. Thus in our work we use the conservative  boost factor of \cite{Moline:2016pbm, Anderson:2010df}
ranging from B=20 for $10^{12} M_\odot$ clusters to B=55 for $10^{15} M_\odot$ clusters  (as opposed to the less conservative one of Bi {\it et al.} \cite{Bi:2006vk} which is roughly a factor of 20 or more larger).

As substructures are to a large extent destroyed by tidal effects near the central parts of the clusters, 
the DM photon annihilation profile will be substantially more extended than the NFW mass profile \cite{Pinzke:2009cp,Pinzke:2011ek,Han:2012uw}. 
We are therefore using the lack of observed line signals in a relatively large region around each cluster to place bounds on the DM contribution. We consider a region of interest which will provide at  least 95 percent of photons from cluster which as mentioned typically corresponds to about a degree on the sky.

\newlinefix
Simulations of \cite{Han:2012au} show that the surface brightness profile of sub-halo emission can be
fitted within $r_{200}$ by
\begin{equation}
J_{sub}(r)=\frac{16 B\mathcal{J}_{NFW}}{\pi
\ln(17)}\frac{D_A^2}{r_{200}^2+16r^2}\;\;\;\;\; (r \leq r_{200}). 
\end{equation}
Following \cite{Han:2012uw}, we take for the subhalo emission surface brightness beyond the
virial radius an exponential decay, 
\begin{equation}
J_{sub}(r)=J_{sub}(r_{200})e^{-2.377(r/r_{200}-1)}\;\;\;\;\; (r \geq
r_{200}). 
\end{equation}

\subsubsection{The Total Differential Flux and Cutoff Radius}

The total annihilation profile, which is the sum of the contributions from a
smooth NFW profile and the subhalo emission, is completely
dominated by subhalo emission except in the very center of the
cluster where the smooth NFW cusp is important. 

Rather than accumulating all the photons emitted by a cluster, 
we perform a signal cut at a radius from the center of the cluster that contains 95\% of the DM signal (from the substructure).  
Thus we take the expected differential flux of photons from a cluster $k$ due to DM annihilation to a line to be
\beq
    \label{eq:J95}
    {d\phi_k \over dE } = \dPhippdE \times(\mathcal{J}_{0.95})_k\ 
\eeq
where 
\begin{equation}
    \mathcal{J}_{0.95}= 0.95 \mathcal{J}_{sub}  \sim 0.95 B  \mathcal{J}_{NFW} 
\end{equation} 
is the integrated J-factor that includes 95\% of the photons from the substructure of the cluster.

Given $ \mathcal{J}_{0.95}$ from theoretical estimates of DM annihilation, we now look for a conical
section from the point of view of the Earth, defined by and angle $\theta_{max}$, that satisfies
\beq
2 \pi  \int_0^{\theta_{max}} J_{sub}  \sin\theta{\rm d}\theta\simeq 2 \pi  \int_0^{\theta_{max}} J_{sub} \theta{\rm d}\theta
=\mathcal{J}_{sub} 0.95=\mathcal{J}_{0.95}\, .
\eeq
This angle then defines a radius 
\begin{equation}
r_{max}=\theta_{max}\times D_A \,\,\,\, {\rm (small} \,\, {\rm angle} \,\, {\rm approximation)}
\end{equation}
that contains 95\% of the DM annihilation signal from substructure (We are not including instrumental error in $r_{max}$. 
We will return to the effect of the point spread function when defining the region of interest).
The angular diameter distance to the cluster $D_A$ is obtained from data as described in the next section.
We will cutoff any signal beyond this radius $r_{max}$, which is typically roughly twice the viral radius of the cluster.
In our analysis, we ignore the smooth component of the halo, thereby underpredicting the DM signal and obtaining slightly more conservative bounds. 
\footnote{
    See dark matter model section (next section) for more comprehensive detail on how $J_{sub}$ is calculated and what it means. 
}

\section{Data Selection}

We perform what is denoted as a stacked cluster approach in our data selection and analysis. 
Previous work done using clusters to constrain dark matter annihilation into a line have taken two different approaches: 
a detailed cluster approach or a stacked cluster approach. 
For a given set of clusters, the detailed approach has stronger bounds than the stacking method.
In the detailed case, each halo is carefully analyzed and finally a joint likelihood is 
performed on tens of objects  \cite{ProkhorovAndChurazov}.
These authors have typically used galaxy clusters which are bright in X-ray (also presumably also dark matter rich).
In the stacking approach, the data is stacked before performing joint likelihood analysis. 
One such stacking approach is \cite{RasatAndLandAndLahavAndAbdalla},
which looks at a cross correlation between photons from the Fermi-LAT with galaxy surveys such as 2MASS and SDSS.  

\subsection{Selecting Galaxy Clusters}

We use Tully's 2MRS catalog of clusters or groups of galaxies,
which is widely regarded as one of the best and most comprehensive catalogs 
of galaxy clusters derived from 2MASS data.  2MRS includes redshift information.
 Tully's catalog of 2MRS data is optimal for our work for two reasons.
First, the large number of clusters guarantees that we will find promising targets from the point of view of large dark matter annihilation signal and no coincidence with background point sources as described further below.
Second, in a proper analysis we would like to have detailed and unbiased statistical description of the halo properties
which will play a part in the dark matter signal, namely the halo's mass and distance.
Some authors have previously exploited the 2MASS data (for example \cite{Xia:2011ax}) 
to construct galaxy cluster catalogs; however, Tully obtains improved estimates of the mass and
distance to the clusters (see \cite{tully} for more details on the various statistical tests used
in his work).

    \begin {table}[h!]
        \label{tab:BrightestGal}
        \caption {Five Brightest  Clusters Used in our Analysis}
        \begin{center}
            \begin{tabular}{c | c |c | c | c | c | c | c | c | c}
            Nest ID                                 &   
            J-factor                            &  
            Brightest (k-band)     &  
            \# galaxies                             &  
            Mass of Cluster                         &       
            G-Lat                                   &   
            G-Lon                                   &  
            Distance                                &  
            Size  
            \\
            &  $\log_{10}(J_\text{nfw}/\GeV^2\text{cm}^{-5})$    & Galaxy In Cluster & &$(10^{12} M_{sum})$ &  (Deg) &  (Deg) & (Mpc) & (Deg) \\\hline
            200002   &17.4   &PGC2801990  &167  &1270   &-7.3   &325    &73 &1.8 \\\hline
            111812   &17.3   &M82         &1    &1.17   &40.6   &141    &4  &3.1 \\\hline
            100128   &17.3   &NGC004594   &15   &93.7   &52.8   &300    &24 &2.2 \\\hline
            101312   &17.3   &NGC5236     &3    &11.4   &31.9   &31     &11 &2.5 \\\hline
            100122   &17.2   &NGC3376     &11   &40.4   &57.7   &233    &19 &2.1 \\\hline
            &&&&&&&& \\
            \end{tabular}
        \end{center}
    \end{table}

We also note that 2MASS data have advantages over other surveys previously used to do studies of DM annihilation from clusters  ---  clusters in X-ray catalogs, SDSS data, and 
using the S-Z effect.    2MASS has the two advantages of providing a larger number of clusters and, since they are nearby ($z<0.01$ ), they have higher luminosities.  For example, the 10 brightest clusters  in our study are nearly three times brighter than the X-ray clusters used in the work of Anderson etal \cite{Anderson:2015dpc}.  Here
brightest means the highest $ \mathcal{J}_{NFW} $.  Table~\ref{tab:BrightestGal} shows the five brightest clusters used in our analysis.
 Indeed 
none of our 16 brightest clusters overlaps with the 16 clusters used in Anderson etal, as shown in  Fig.~\ref{700BGal} in comparison with Fig.~\ref{600BGal}.  
One reason for the lack of overlap is that 2MASS contains so many clusters that we simply 
throw away those that overlap with any of the 150 brightest objects in the Fermi/LAT point 
source catalog.  Whereas two of the brightest 
objects in the sky from the point of view of DM annihilation are Virgo and Fornax, 
we do not include them in our studies due to the point source contamination.  Anderson etal
dealt with the point sources by doing careful modeling of the clusters and cutting out the point
sources using spatial information.  We, on the other hand, throw them out completely in favor
of using some of the many other clusters in the 2MASS data. In principle it would be interesting to 
add them back in to obtain tighter bounds on DM annihilation from the combination of the data.

Henceforth our studies are performed for three different cuts on halos:  all the halos in the Tully 2MASS catalog, the 2500 brightest halos, and the 600 brightest halos. Again, we use the word ``brightest" to mean those clusters with the highest value of $ \mathcal{J}_{NFW} $.  Table~\ref{tab:TotalJ} illustrates the total value of 
$ \mathcal{J}_{NFW} $ added up for all the clusters in each of the three cuts.

  \begin {table}[h!]
                  \caption { Total value of 
$ \mathcal{J}_{NFW} $ added up for all the clusters in each of the three cuts}  
        \begin{center}
            \begin{tabular}{l | c } 
            Cut:                        & $\mathcal{J}_{NFW}$ Total  $(GeV)^{2} / (cm^{5})$     \\  \hline
            All Clusters                & $9 \times 10^{19}$                          \\  \hline
            2500 Brightest Clusters     & $5 \times 10^{19}$                          \\  \hline
            600 Brightest Clusters      & $3 \times 10^{19}$                          \\  \hline
            \\
              
            \end{tabular}
	
        \end{center}
        \label{tab:TotalJ}
    \end{table}

 \newpage

\begin{figure*}[t!]
    \label{fig:ClustersAll}
    \includegraphics[width=\textwidth, keepaspectratio]{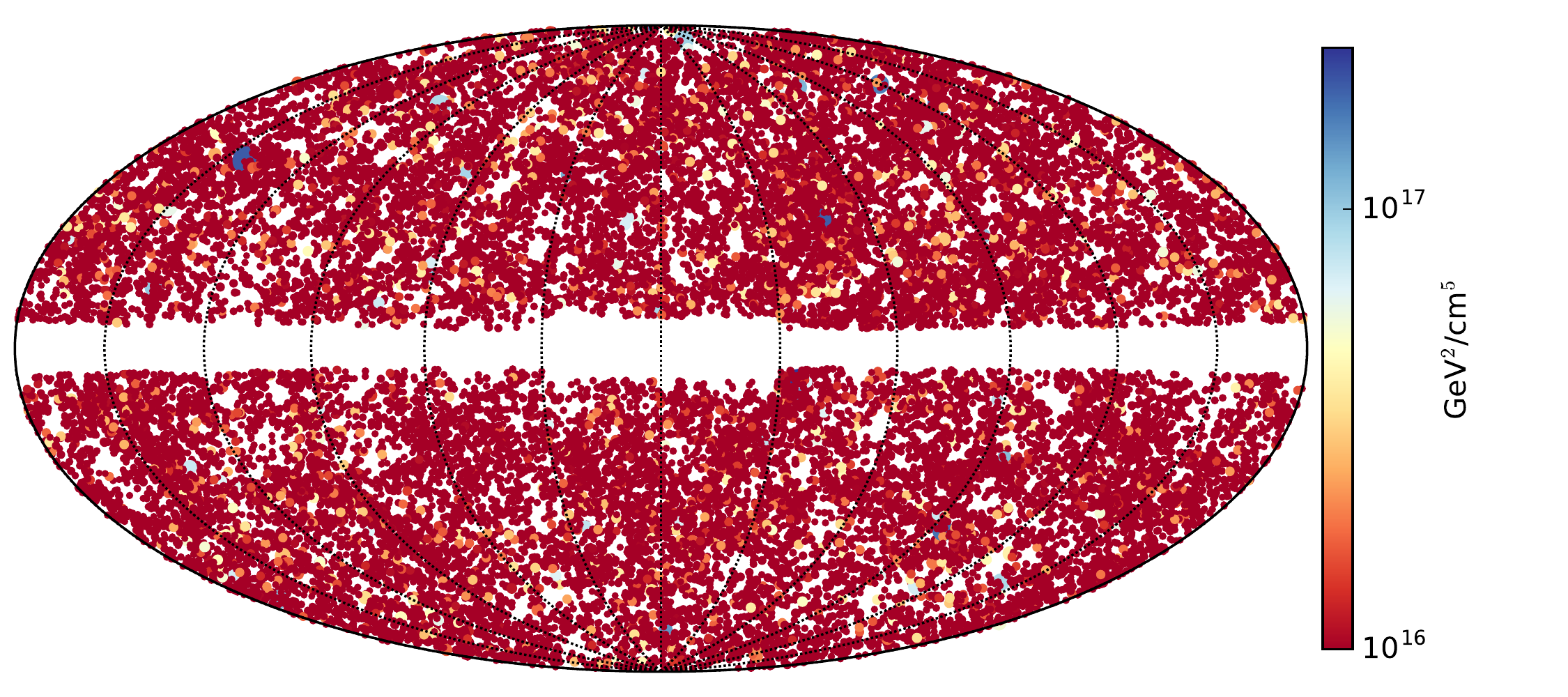}
    \caption{All clusters from Tully 2MASS Groups Catalog. We plot the position and size of the galaxy with a circle. The radius of the circle gives the Region of Interest for each galaxy.
    The color coding denotes the galaxy's $J_\text{NFW}$.}
    \label{AllCluster}
\end{figure*}

\newlinefix 
\begin{figure*}[t!]
    \includegraphics[width=\textwidth, keepaspectratio]{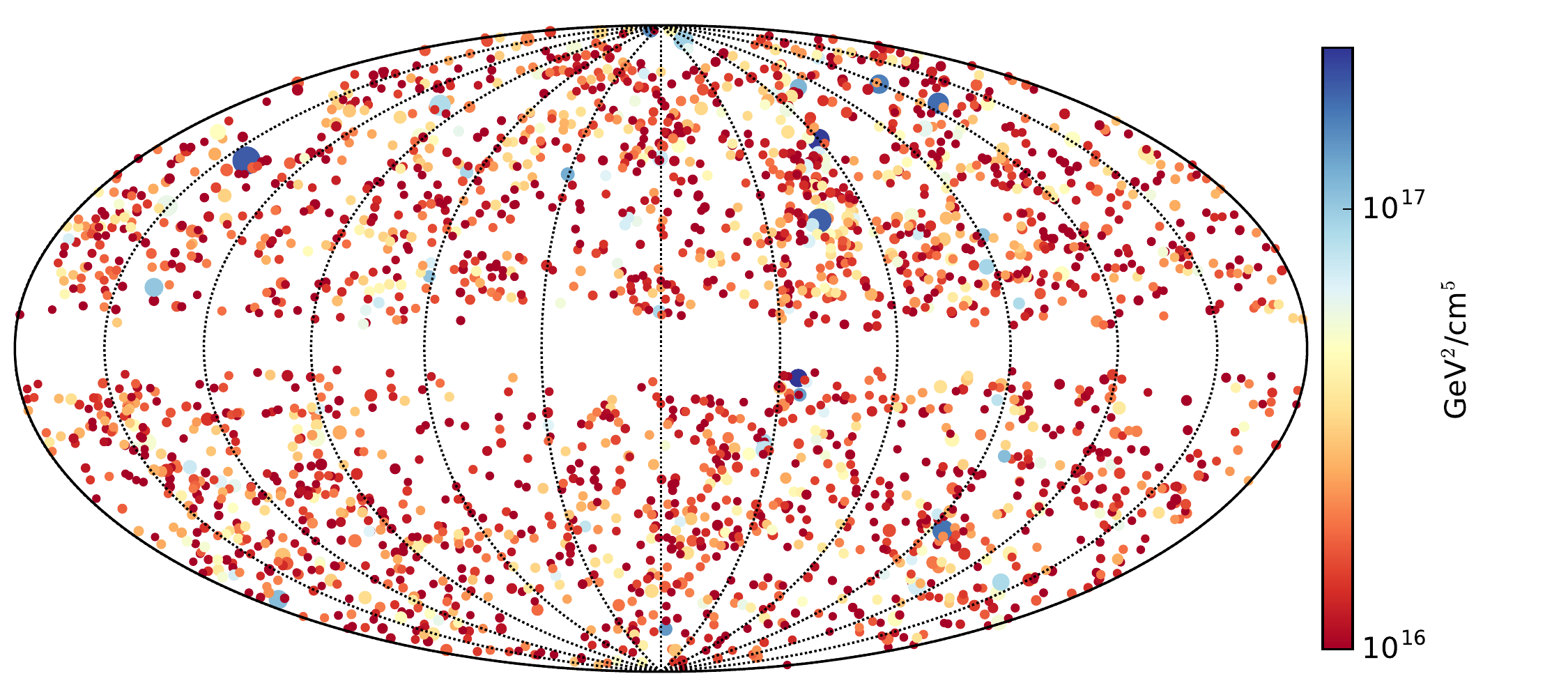}
    \caption{Brightest 2500 clusters from Tully 2MASS Groups Catalog. Same symbols and color coding
    as in  Fig.\ref{AllCluster}.}
\end{figure*}

\newlinefix 
\begin{figure*}[t!]
    \includegraphics[width=\textwidth, keepaspectratio]{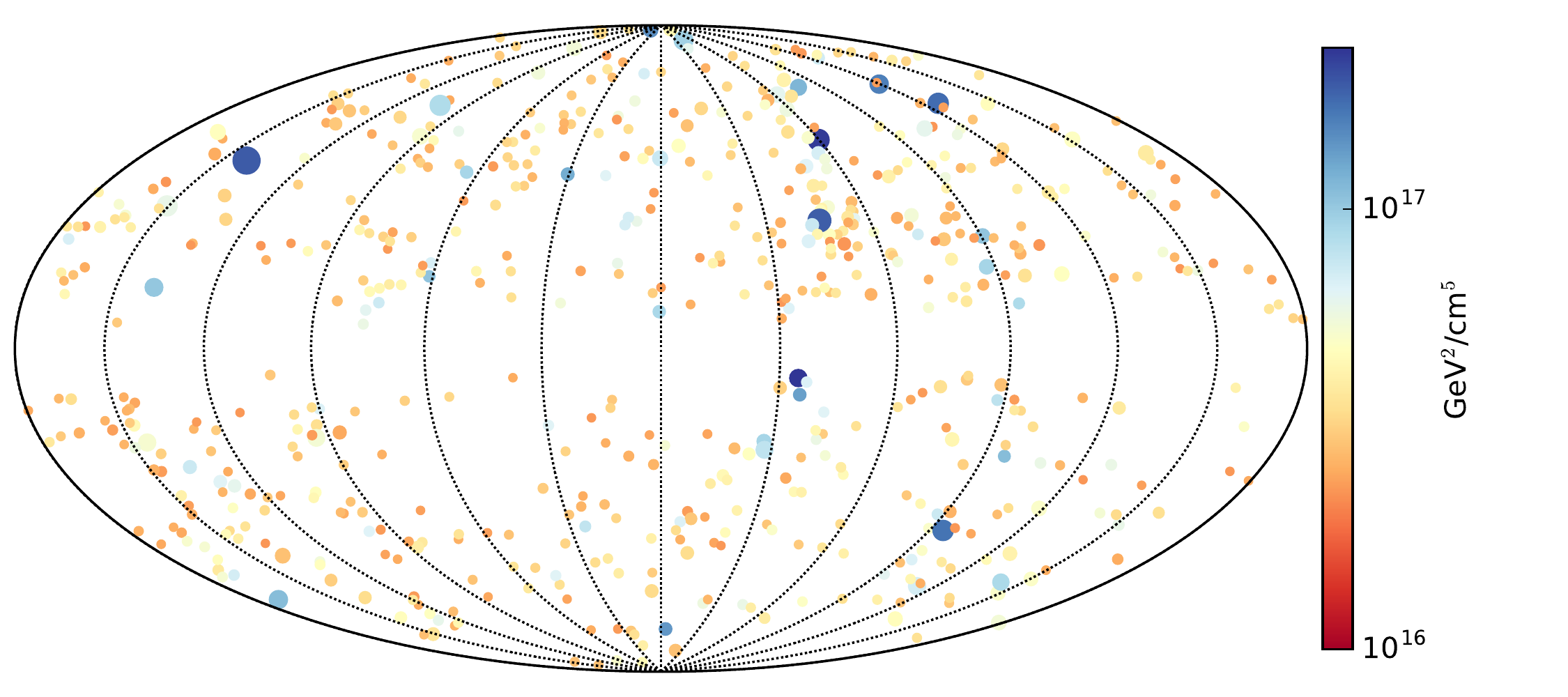}
    \caption{Brightest 600 clusters from Tully 2MASS Groups Catalog.  The galaxy clusters have a $J_\text{NFW}$ larger than $2.2\times10^{16}$ (GeV$^2$cm$^{-5}$).
Same symbols and color coding
    as in  Fig.\ref{AllCluster}.}
    \label{600BGal}
\end{figure*}

\newlinefix 
\begin{figure*}[t!]
    \includegraphics[width=\textwidth, keepaspectratio]{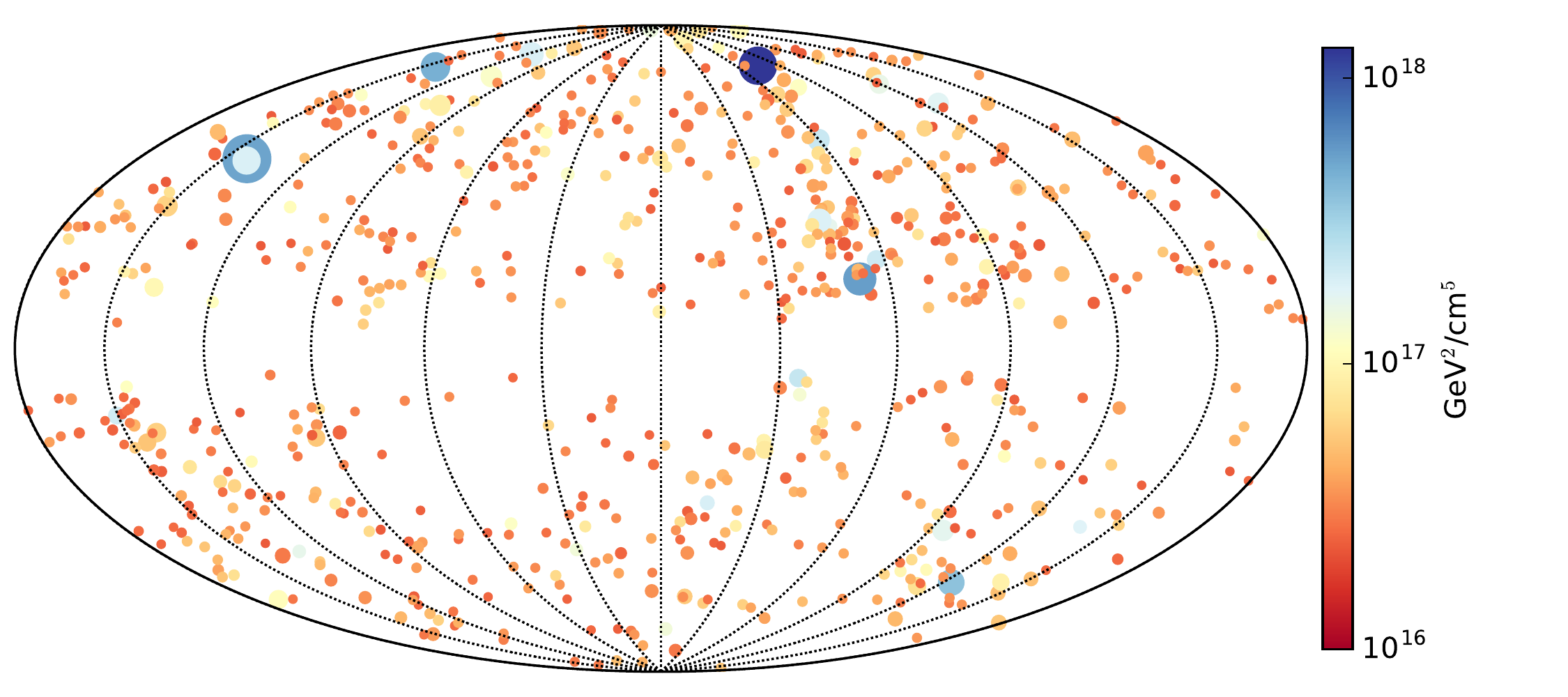}
    \caption{As a comparison with Fig.\ref{600BGal}, we consider the case where we also include clusters which  overlap with
    one of the 150 point sources. With the same cut off as in Fig.~\ref{600BGal}, we now have 100 more galaxy clusters. Note that 
    the brightest clusters are Virgo and Fornax in that order. In our analysis, we do not include Virgo and Fornax since they overlap
    with one of the bright 150 point sources.}
    \label{700BGal}
\end{figure*}

\subsection{Selecting $\gamma-rays$ from Fermi-LAT }

The LAT instrument aboard the Fermi satellite is a pair-conversion telescope measuring $\gamma$-rays in the energy range from 20 MeV to $>$ 500 GeV. For a more detailed description, see~\cite{Atwood:2009ez,Ackermann:2012kna}. 
We analyze five years of public Fermi-LAT Pass 7 reprocessed data taken between  4 Aug 2008 to 8 Mar 2012 in 
the energy range between 20 and 500 GeV.

We apply the celestial zenith-angle cut $ \theta < 100 ^{\circ} $ in order to avoid contamination with the earth albedo, as well as the recommended quality-filter cut $DATA\_QUAL==1$.
We use the $ULTRACLEAN$ events selection, i.e. the highest quality data, from $P7REP\_ULTRACLEAN\_V15$.
We use both front- and back-converted events. 
The selection of events as well as the calculation of exposure maps is performed using the most recent version of Fermi Science Tools as of 2015 Jul 1.\footnote{See \url{http://fermi.gsfc.nasa.gov/ssc/data/analysis} }
We now select the subset of the photons in the Fermi-LAT data coming from the direction of the clusters in all of our catalogs.

\begin{figure*}[t!]
    \label{fig:allPhotons}
    \includegraphics[width=\textwidth, keepaspectratio]{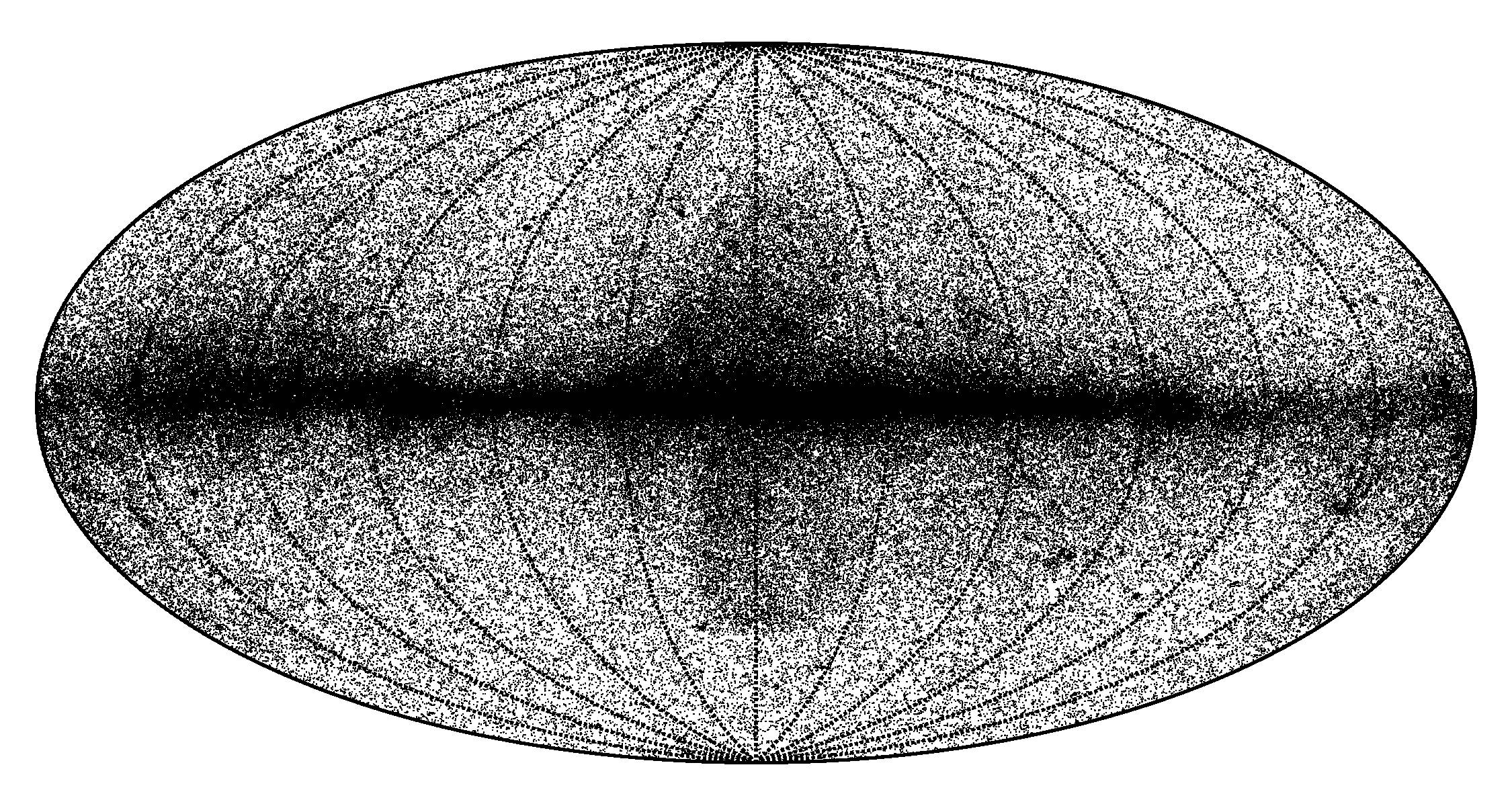}
    \caption{Fermi-LAT photons from ULTRACLEAN $P7REP\_ULTRACLEAN\_V15$ dataset}
\end{figure*}

As mentioned previously, we sub-select again to reject background by removing the 5\% brightest point sources identified by Fermi-LAT which are coincident with the clusters we are studying.  Consequently we only include clusters which do not overlap with any of brightest 150 Fermi-LAT point sources.  
By doing so, we exclude some of the brightest clusters (from the point of view of $\gamma$-ray production due to DM annihilation) on the sky such as Fornax and Virgo, see Fig.~\ref{600BGal}.
Still after this sub-selection, we retain the vast majority of the clusters in the catalogs included; 
on the order of a few hundred clusters were removed out of 26,000 we work with. 
Similarly, when we apply the cut down to only the 600 brightest clusters, we have lost about 100 clusters. 
Compare Fig.~\ref{600BGal} with Fig.~\ref{700BGal}.

\begin{figure*}[t!]
    \includegraphics[width=\textwidth, keepaspectratio]{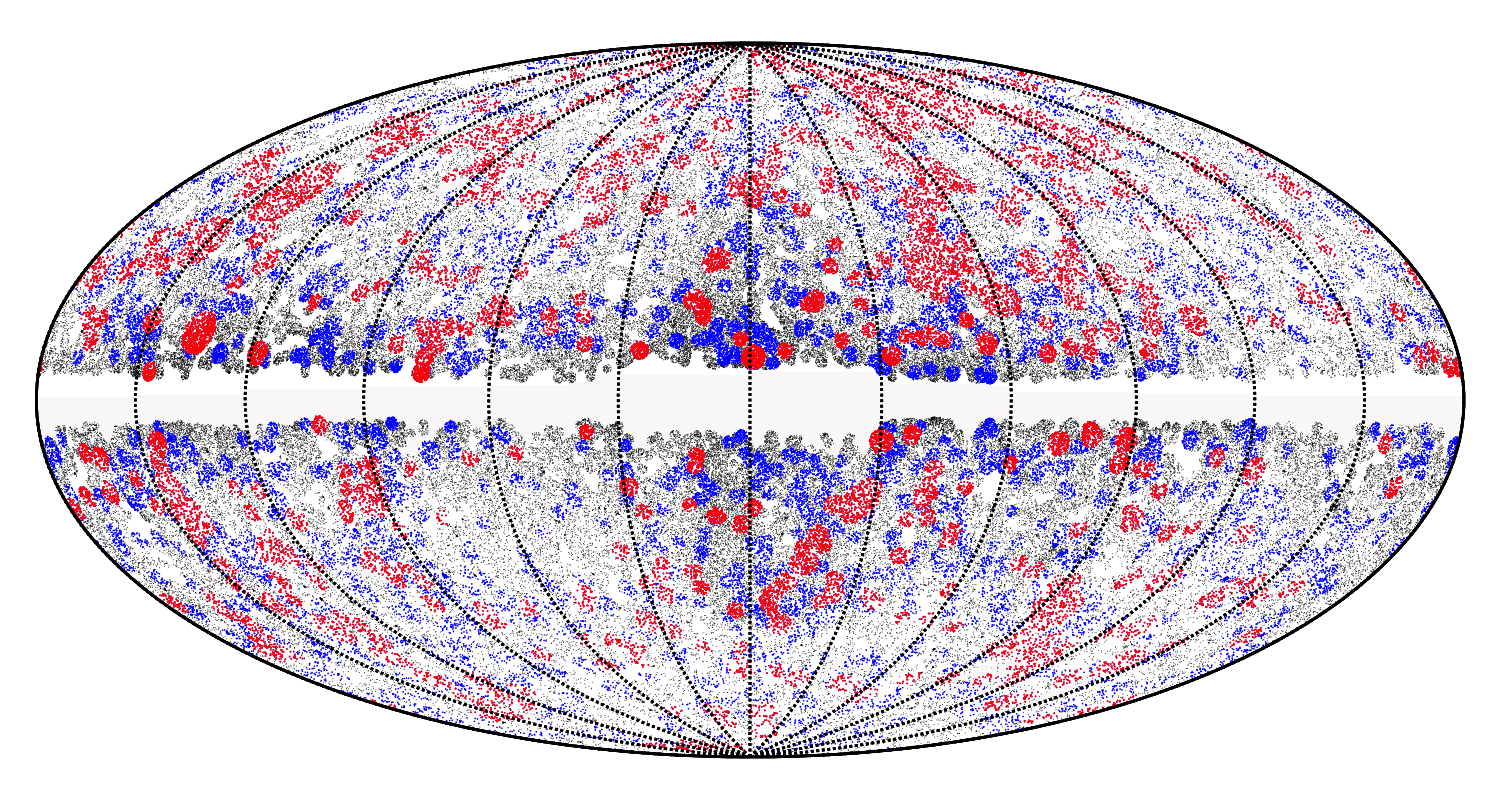}
    \caption{
         Fermi-LAT photons from ULTRACLEAN $P7REP\_ULTRACLEAN\_V15$ associated with Galaxy Clusters. 
         Photons marked with red are associated with the brightest 600 clusters, blue with the brightest 2500 clusters, and grey with any cluster.
        }
         \label{fig:PhotonsWithHalos}
\end{figure*}

\subsubsection{Region of Interest}

To compare the expected number of photons to observations, we must determine which photons in the Fermi-LAT data to include in our analysis of a given cluster. There are two ingredients in this choice. Previously we found 
(see Eqn. (18)) the maximum radius $r_{max}$ from the center of the cluster that satisfies the requirement of retaining 95\% of the signal from the substructure. Second, we must account for the instrumental smearing of the directionality of the signal due to the point spread function (PSF). 
We note that the Fermi-LAT PSF varies as a function of energy, and above energies of 40 GeV the PSF varies slowly.  

To calculate the selection radius of photons for a given cluster, 
the most accurate approach would be to convolve the point-spread function (PSF) of Fermi-LAT with the DM profile of the cluster.
We can, however,  take a simpler approximation.
After extracting the PSF from the Fermi Tools, for each value of the DM mass, we calculated the radius within which 95\% of the photons lie due to the uncertainty of the detector $r_{PSF,0.95}$.  Specifically,
the 95 percent PSF containment angle is radial angle describing a circle around a point source 
such that 95 percent of the photons from that point source 
are observed in that angle accounting for instrument error. 
The PSF containment angle typically ends up being on the order of a degree for energies above 100 GeV.  
Then for each cluster, we consider a Region of Interest (ROI) for incoming photons taken to be the sum 
\begin{equation}
r_{ROI} = r_{max} + r_{PSF,0.95} \, .
\end{equation}
Our region of interest corresponds to a radius within which we look in the Fermi-LAT data for photons from any given cluster in the Tully data.

\begin{figure}
    \includegraphics[width=\textwidth, keepaspectratio]{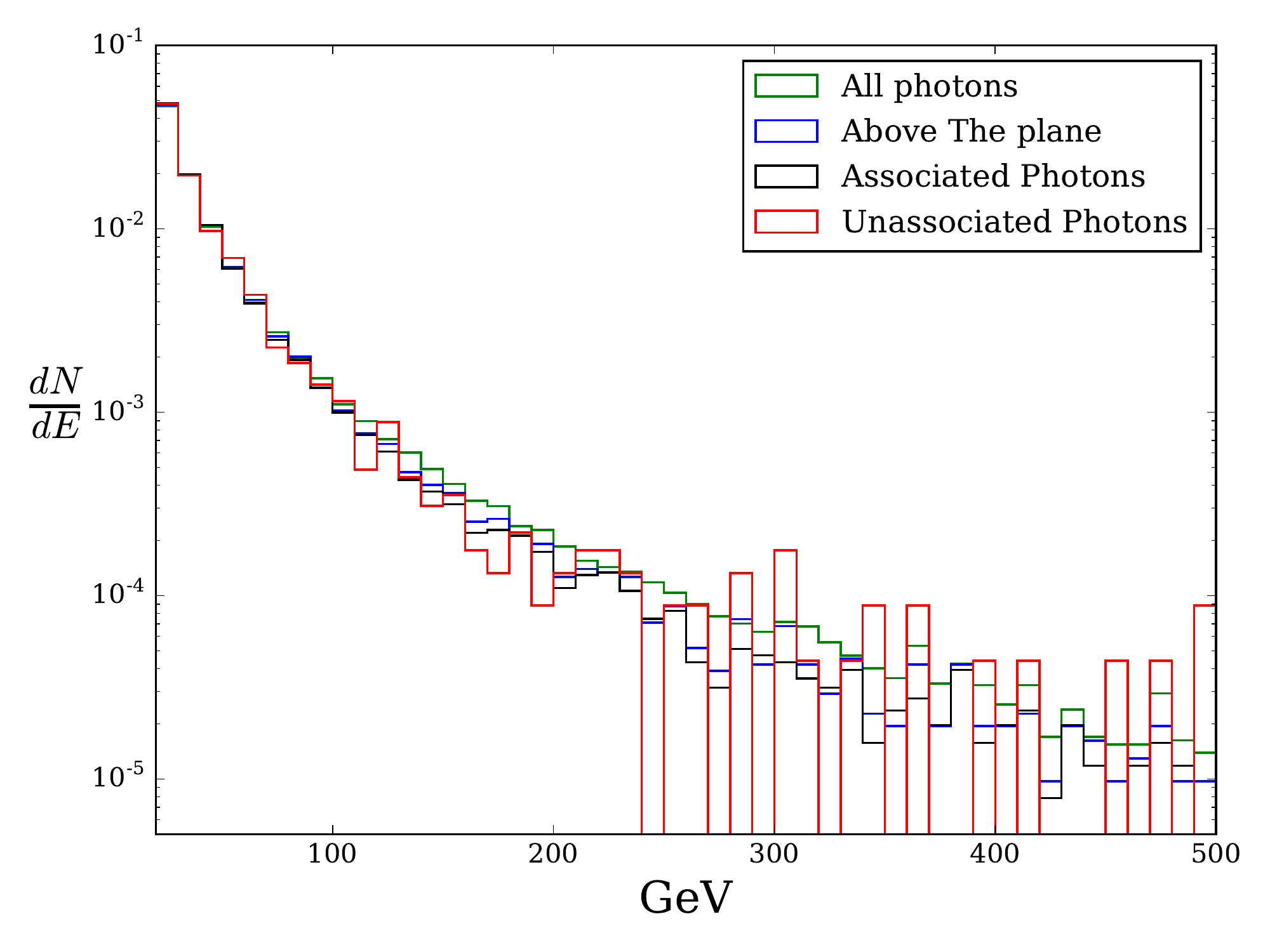}
    \caption{
    Binned energy spectra for Fermi-LAT $\gamma$-rays. All curves are normalized such that the area under the curve integrates to 1.
    The green curve plots the energy spectra for all the Fermi-LAT observed photons. 
    Black is for photons from outside the Galactic Plane( $|b| > 10\degree$) that are associated with galaxy clusters.
    Red is for photons again outside the Galactic Plane but unassociated with galaxy clusters and unassociated with any of  the 150 brightest point sources.
    Blue is for all photons outside the Galactic Plane (the sum of associated and unassociated). 
    Energy values run from 20 GeV
    to 500 GeV in bins of 10 GeV.
    }
        \label{fig:SpectrumComparison}
\end{figure}

\begin{figure}
    \includegraphics[width=\textwidth, keepaspectratio]{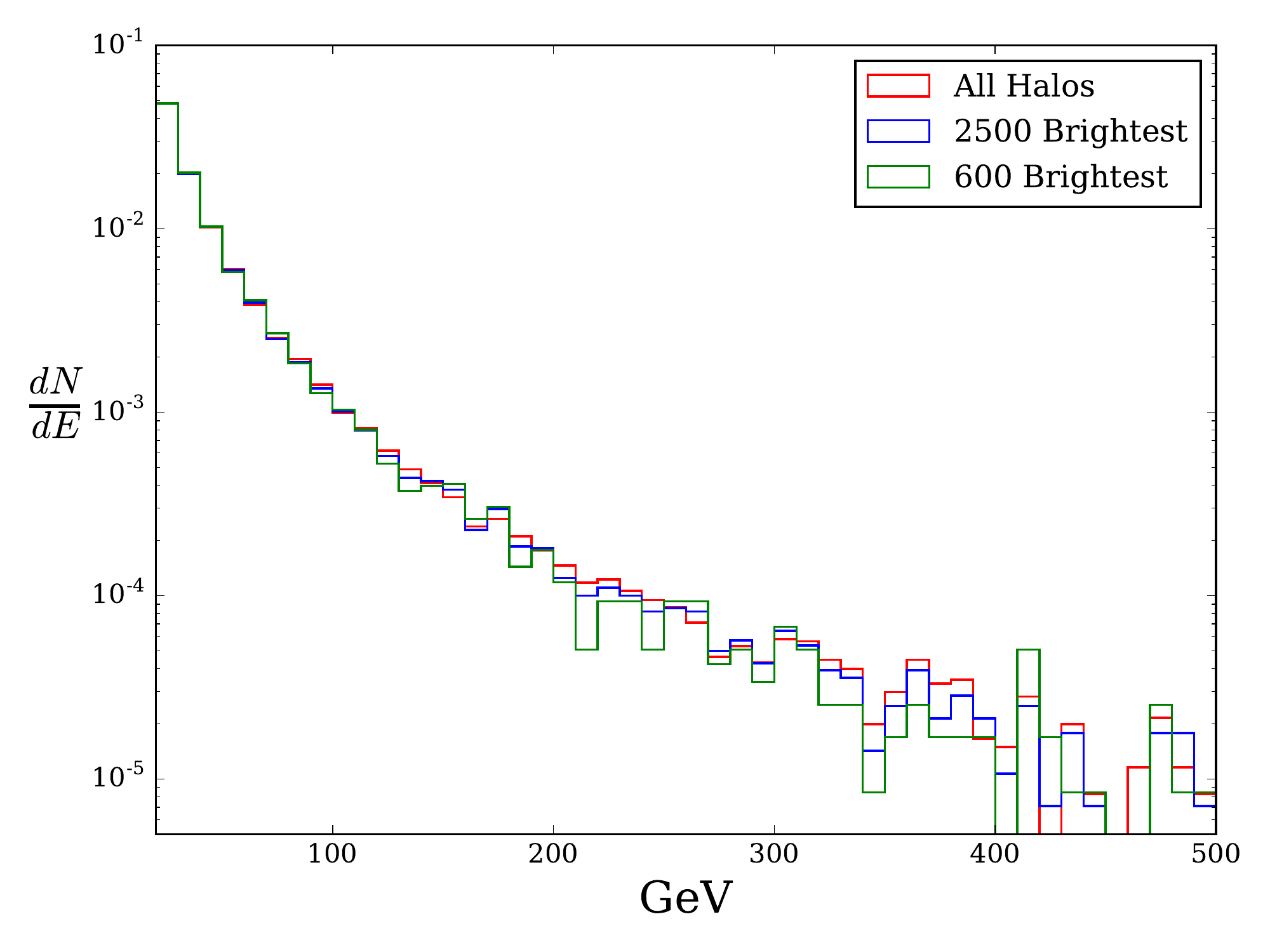}
    \caption{
    Binned energy spectrum of Fermi-Lat $\gamma$-rays associated with: all the clusters, the 2500 brightest clusters, and the 600 brightest clusters shown from 20 GeV-500 GeV in 10 GeV bins.  Again,
    all curves are normalized such that the area under the curve integrates to 1.
    }
     \label{fig:SpectrumHalos}
\end{figure}

\subsection{ Photon Number Count and Energy Spectrum}

The photon number count is shown in Table~\ref{tab:photons} for a variety of different selection cuts for photon
energies in the range 10 GeV $< E_\gamma <$ 1 TeV. Only photons from outside the Galactic Plane with 
Galactic latitude $|b| > 10\degree$ are included in all cases.  The total photon count outside the Plane is shown,
followed by photons associated with all clusters, photons associated with the 2500 brightest clusters, the 600
brightest clusters, and the 150 brightest point sources in the Fermi-LAT data.  Finally, the photon number
count from outside the Galactic Plane but unassociated with clusters is shown as well.

    \begin {table}

        \caption { Number of Photons for Different Selection Cuts with $ 10 {\rm GeV} < E_{\gamma} < 1 {\rm TeV} $}
        \begin{center}
            \begin{tabular}{l | c }
            Selection Cut  & Total Photon Count \\ \hline
            Outside Galactic Plane   & 89,000 \\ \hline
            Associated With All Clusters                & 75,000 \\  \hline
            Associated With 2500 Brightest Clusters     & 35,000 \\  \hline
            Associated With 600 Brightest Clusters      & 12,000 \\  \hline
            Associated With 150 Brightest Point Sources &  7,000 \\  \hline
            Unassociated                                &  7,000
            \\
            \end{tabular}
        \end{center}
                \label{tab:photons}
    \end{table}

Fig.~\ref{fig:SpectrumComparison} illustrates the photon energy spectra for a variety of selection cuts
similar to those in the Table.  The green curve illustrates the spectrum 
 for all the Fermi-LAT observed photons. 
    Black is for photons from outside the Galactic Plane( $|b| > 10\degree$) that are associated with galaxy clusters.
    Blue is for all photons outside the Galactic Plane (the sum of associated and unassociated). 
    Red is for photons again outside the Galactic Plane but unassociated with galaxy clusters and unassociated with any of  the 150 brightest point sources.
    We have normalized
the curves such that the area under each curve will equal one. All of the curves have similar shapes
except that  the green curve appears to fall less steeply with energy, because it includes
 sources in the plane of the Galaxy. The large error bars in the unassociated spectrum (red curve) at high energy are due to the low number of photons in that case (as shown in Table~\ref{tab:photons}).

Fig.~\ref{fig:SpectrumComparison} can be used to illustrate two points.  First, within our  energy 
windows, the spectra can roughly be fit by power laws.  Thus it is reasonable to use a series of 
power laws (within each energy windows) to approximate the background. The unassociated photons in particular are a separate sample
from the clusters we are studying and provide a reasonable background estimate.  Our maximum likelihood approach to handle background will be described in the next section.
Second, a line signal would show up as a bump in this figure and, at least by eye, there doesn't appear to be 
any.  Our analysis will confirm that no line is in the data at any great significance, hence allowing us to 
place limits on DM properties.

We also illustrate how the spectrum will vary for our different cuts on the clusters and galaxies in Figure~\ref{fig:SpectrumHalos}.  We plot
the three different case considered in the paper: the spectrum for all of the clusters,
the spectrum from 2500 clusters, and for 600 clusters. The spectral shape appears to be roughly
the same for the different cases, yet with fluctuations between the different data sets that will 
affect our resulting bounds.

\section{Expected Number of Photons from Signal and Background in the Fermi-LAT}

 We follow the approach of Weniger \cite{weniger_line}, Section 2.5
to calculate the total number of expected  events $\nu_w$ in the energy window $(w)$ between $E_w^-$\ldots$E_w^+$
\begin{equation}\label{Vw1}
  \nu_w =\sum_k^{\text{clusters}}\int d\Omega_k \int_{E_w^-}^{E_w^+} dE 
  \int dE' \int_0^\pi d\theta \sum_{j=f,b} D(E|E', \theta, j) A(E', \theta, j)
  \bigg(\frac{dT}{d\theta}\bigg)_k \mathbf{I}(E')_k\;.
\end{equation}
where we have summed over all of the clusters $k$ and integrated over the solid angle $d\Omega_k$ over each cluster.

Here we have defined:
\begin{equation}
    \begin{aligned}
        \theta              &= \text{ impact angle of photon events with respect to the instrument axis,  } \\
        (dT/d\theta)_k      &= \text{ the observational time at this impact angle for a cluster $k$,  } \\
        A(E', \theta, j)    &= \text{ the effective area, } \\
        E'                  &= \text{ the true energy, } \\
        j=f,b               &= \text{ denotes front- and back-converted events, } \\
        E                   &= \text{ the reconstructed energy } \\
        D(E|E',\theta, j)   &= \text{ the energy dispersion of the LAT. } \\
    \end{aligned}
\end{equation}

We have also defined the following flux quantities:
\begin{equation}
    \begin{aligned}
    (I_{m_{\chi}})_{k}  &= \text{differential flux from dark matter into a line (see Eq. \ref{eqn:dphidE} ) } \\
    (I_{b})_{k}         &= \text{differential background flux } \\
    I_{k}               &= (I_{m_{\chi}})_{k} + (I_{b})_{k} \text{=  total differential flux}
    \end{aligned}
\end{equation} 

\begin{figure*}[t!]
      \includegraphics[width=\textwidth, keepaspectratio]{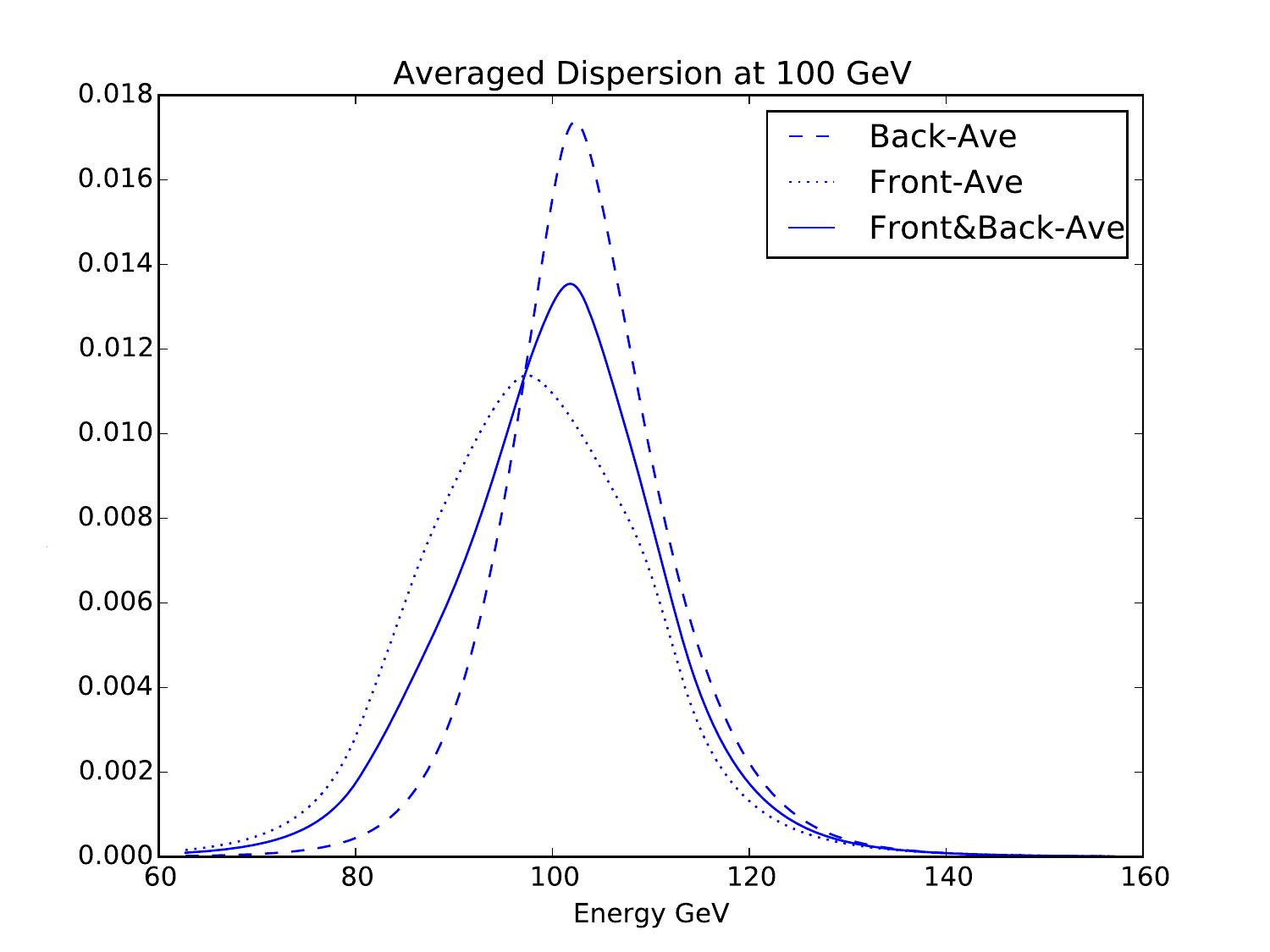}
    \caption{ 
        The average effective energy dispersion function  for the case of 100 GeV photon energy (see text for details).  The average dispersion associated with front- and back-converted events is shown separately as dotted and dashed lines respectively. The solid line plots the average of the two, $D_{eff}(E | E'=100GeV )_{\text{ave}}$ in Eq.(\ref{eq:Deffave}).}
         \label{fig:AverDisp}
\end{figure*}

For a given dark matter mass, we picked a window which is sufficiently large to effectively model the background.
The energy resolution of Fermi-LAT is roughly 10-20 percent over the energies we are considering
(see Fig.~\ref{fig:AverDisp} ).
We take the energy window to be $m_\chi/1.4 < E < 1.4 m_\chi$, but with a maximum value for the lower
limit of the energy window $E^-_{max} =  200$GeV
(due to a dearth of very high energy photons). 
With a 100 GeV line, our window would then go from 71 to 140 GeV. 
We have found that taking a 
larger or slight smaller window does not affect
the overall constraint, as has been similarly shown in \cite{weniger_line}.

\newlinefix
\indent We average over the instrument angle and back-and-front converted events.  
Again we follow Weniger and define an effective exposure for a cluster $k$ as 
\begin{equation}
\label{eq:xeff}
    X_{eff}(E')_k \equiv \int_0^\pi d\theta \sum_{j=f,b} A(E', \theta, j) \bigg(\frac{dT}{d\theta}\bigg)_k\;
\end{equation}
and an effective energy dispersion for cluster $k$ as
\begin{equation}
    \label{eq:Deff}
    D_{eff}(E|E')_k\equiv \int_0^\pi d\theta \sum_{j=f,b} D(E|E', \theta, j) P(\theta, j,
    E')_k\;.
\end{equation}

\newlinefix 

For cluster $k$, the expected distribution of the conditional parameters $\theta$ and
$j$ for events with energy $E'$ is
\begin{equation}
    P(\theta, j| E')_k\equiv\frac{A(E', \theta, j) (dT/d\theta)_k}{X_{eff}(E')_k}\;.
\end{equation}
From the publicly available Fermi Science Tools\footnote{See \url{http://fermi.gsfc.nasa.gov/ssc/data} }, we obtained the dispersion function $D(E|E', \theta, j)$, effective area 
$A(E', \theta, j)$ , and $dT/d\theta$ for each cluster. 

Now that we have obtained the effective energy dispersion function for each cluster in Eq.(\ref{eq:Deff}), we can obtain an ``average effective energy dispersion function"  $ D_{eff}(E|E')_\text{ave}$
over the entire sky.   We first average over the clusters to find 
($dT/d \theta)_{ave}$. With this new quantity,  we then
average over the instrument angle and back-and-front converted events
to obtain the average effective dispersion function over the clusters:
\beq
 P_{eff}(\theta, j| E')_\ave\equiv\bigg( \frac{A(E', \theta, j)( dT/d\theta)_\text{ave}}{(X_{eff}(E'))_\text{ave}}\bigg)\;
\eeq
where
\begin{equation}
    X_{eff}(E')_\text{ave} \equiv \int_0^\pi d\theta \sum_{j=f,b} A(E', \theta, j)_k \bigg(\frac{dT}{d\theta}\bigg)_\text{ave}\;.
\end{equation}
and
\begin{equation}
    \label{eq:Deffave}
    D_{eff}(E|E')_\text{ave}\equiv \int_0^\pi d\theta \sum_{j=f,b} D(E|E', \theta, j) P(\theta, j,
    E')_\text{ave}\; .
\end{equation}
In  Fig.~\ref{fig:AverDisp}, we show $D_{eff}(E|E')_\text{ave}$ for the case ($E'=m_\chi=100$ GeV).
Weniger argues that the average effective dispersion function is a good approximation at any location in the sky.
Thus, for a given dark mass $m_\chi$, for all of the clusters we will use the same average dispersion function.

We now calculate the expected number of photons in some energy window $(w)$,
\begin{equation}
  \nu_w = \int_{E_w^-}^{E_w^+} dE \int dE'
  D_{eff}(E|E')_\ave \sum_k^{\text{clusters}}X_{eff}(E')_k\int d\Omega_k   \mathbf{I}(E')_k\;.
\end{equation}
Here the solid angle integral $d\Omega_k$  is over the ROI for the given cluster.
The effective exposure varies slowly across the sky; we note that we do not
use the average effective exposure here, but instead the local one.  Since the ROI
of a halo is typically only of order a degree  on the sky, we can treat the exposure as constant across a cluster and
take it out of the integral over  $d\Omega_k$.

We take the differential photon number count due to the background component to be a power law
(within an energy window ranging from $E_w^-=m_\chi/1.4$ to $E_w^+=1.4m_\chi$) of
the form
\beq
  {d\nu_b \over dE} =
\int dE'
 D_{eff}(E|E')_\ave \sum_k^{\text{clusters}} X_{eff}(E')_k \int d\Omega_k  \mathbf{I}_\text{b}(E')_k
  =\beta\bigg(\frac{E}{m_\chi}\bigg)^{-\gamma} .
  \eeq
 Different WIMP masses have different power laws for their background fits. 
 
Now for the DM signal, the differential photon number count within the same energy window
would be  
\beq
 {d\nu_{m_\chi} \over dE} =
\int dE'
  D_{eff}(E|E')_\ave \sum_k^{\text{clusters}} X_{eff}(E')_k \int d\Omega_k  \mathbf{I}_{m_\chi}(E')_k= 
  D_{eff}(E|m_\chi)_\ave\times \frac{\langle\sigma v\rangle}{4\pi m_\chi^2}\times \mathcal{(JX)}_T \, .
  \eeq
  The first term on the RHS accounts for the instrument and the second term accounts for
 the particle physics. The last term defines a new quantity (which we call the exposed integrated J-factor) that accounts for both the astrophysics and exposure for each cluster,
 \beq
 \label{eq:J0}
 \sum_k^{\text{clusters}} (\mathcal{J}_{0.95})_k X_{eff}(E_\gamma)_k =\sum_k^{\text{clusters}}\mathcal{J}_k
 \mathcal{X}_k=\sum_k^{\text{clusters}}\mathcal{(JX)}_k=\mathcal{(JX)}_T\, .
  \eeq
\smallskip
Here we have used Eq. (\ref{eq:J95}) to determine the integrated J-factor $(\mathcal{J}_{0.95})_k$ for cluster k.
The notation $\mathcal{(JX)}_T$ is intended to guide the eye of the reader to the fact that this quantity
has been obtained by summing the product of two quantities: the integrated J-factor for each cluster times
its exposure. The subscript $T$ refers to the fact that this a total quantity. In addition, we use 
$\mathcal{J}_k\mathcal{X}_k=\mathcal{(JX)}_k$ as a 
shorthand for the individual J-factor and exposure for a given galaxy $k$, which will be useful in describing the 
J-factor Likelihood.

We now have our expected  photon energy spectrum,
\beq
\label{eq:specTotal}
\frac{d\nu_w}{dE}=\frac{\langle \sigma v\rangle}{4\pi m_\chi^2} \times D_{eff}(E|m_\chi)_\ave\times \mathcal{(JX)}_T+\beta\bigg(\frac{E}{m_\chi}\bigg)^{-\gamma}\, .
\eeq
  The first term gives the total flux from dark matter annihilation from all of the clusters used in our analysis.
The second term account for the observed background.

\section{Likelihood Analysis}

We performed an un-binned analysis with our likelihood given as
\beq
\like(m_\chi, \langle\sigma v\rangle,\beta,\gamma,\mathcal{JX})=\bigg(\frac{\nu_w^N}{N!}e^{-\nu_w}\bigg)
\times\prod_j^N \frac{1}{\nu_w}\frac{dE_j}{dE}\times\like_{\mathcal{JX}}
\eeq
where $N$ is the total number of observed photons 
in the energy window $w$ between $E_w^-$ and $E_w^+$,
$\nu_w$ is the number of expected photons in the window $w$, 
$E_j$ is the observed energy of a photon, 
$dE_j/dE$ is given by Eq. (\ref{eq:specTotal}).
The product is over all photons in $w$ for a given DM mass $m_\chi$. 
The first term on the RHS (a Poisson distribution) 
gives the probability of observing N photons with an expectation of $\nu_w$ photons.  
The second term accounts for the probability of observing any one of the $N$ photons.  
The last term ($\like_{\mathcal{JX}}$) is the likelihood function
for the exposed integrated J-factor given in Eq. (\ref{eq:J0}), which we now derive below.

\subsection{ J-Factor Likelihood Function}

We take a Gaussian probability distribution
for the Exposed Integrated J-factor likelihood function
\beq 
\like_\mathcal{JX}=\frac{1}{\sigma_T\sqrt{2\pi}}{\rm exp}-\bigg(\frac{(\mathcal{JX})-(\mathcal{JX})_T}{\sigma_T}\bigg)^2\, .
\eeq
where $\sigma_T$ incorporates the individual errors on the exposed integrated J-factor for each cluster and any other 
systematic errors. $T$ refers to total as in total error for $\sigma_T$. $(\mathcal{JX})$ is the free variable and $(\mathcal{JX})_T$ is the expected integrated J-factor given 
in Eq.(\ref{eq:J0}).

Formally the relative error for the exposed integrated J-factor for a cluster will go like
\beq
\frac{\delta(\mathcal{JX})_k}{(\mathcal{JX})_k}=\frac{\delta\mathcal{J}_k}{\mathcal{J}_k}
\eeq	
where $\delta$ refers to the error for a given quantity.
We have ignored the errors associated with the galaxy's exposure $\mathcal{X}_k$.
We note that the J-factor for a given cluster goes like
\beq
\label{J(M)}
J\sim \frac{M^{0.27}}{D^2}\sim \frac{M^{0.27}V^2}{H_0^2}
\eeq
which is derivable from Eq.~(\ref{J_NFW}) by simply rewriting $r_s$ and $\rho_s$ in terms of the mass of the halo ($M$).
$D$ is the distance to the halo. The second step follows from the Hubble law, where $V$ is the velocity of the galaxy and H$_0$
is the Hubble constant. In deriving the above expression,
we have neglected two effects: one- the dependence of the denominator of $\rho_s$  on $M$ and
 two- the mass dependence of the boost
factor on $M$. We have two arguments to justify our simplification. First, the boost factor is a very slowly varying function of $M$. Also
the denominator is very weakly dependent upon mass (a very small fractional power).
Secondly, error due to velocity and error due to the Hubble constant are larger. 

We first provide an expression of errors for  galaxy k with
\beq
\label{galaxyError}
\sigma_k^2=(\mathcal{JX})_k^2\bigg(\frac{\delta\mathcal{J}_k}{\mathcal{J}_k}\bigg)^2=
(\mathcal{JX})_k^2\bigg[\bigg(0.27\bigg(\frac{\delta M}{M}\bigg)_k\bigg)^2+\bigg(2\bigg(\frac{\delta V}{V}\bigg)_k\bigg)^2\bigg].
\eeq
We add the relative errors in mass $\delta M/M$ and in velocity $\delta V/V$ in quadrature since they are independent.
We note that for most of the clusters 
$(\delta M/M)_k\sim0.20$ and $(\delta V/V)_k\sim0.10$, which we will justify below.
Including the uncertainty in the Hubble constant, we find the total error,
\beq
\label{TotalError}
\sigma_T^2=\sum^N_k \sigma^2_k+\bigg(2(\mathcal{JX})_T\frac{\delta \text{H}_0}{\text{H}_0}\bigg)^2
\eeq
where we sum over all $N$ of the clusters in our analysis.
 For smaller (larger) values of H$_0$, we can see from Eq. (\ref{J(M)}) that the J-factors of the clusters increase (or decrease), so that they appear brighter
 (dimmer).  Again, we add the error in H$_0$ in quadrature.
The errors in velocity and halo mass are independent of the Hubble constant.
 
  In our code we numerically calculate Eq.~(\ref{TotalError}).  We find that the errors associated with $H_0$  dominate in $\sigma_T$
  over those related to the properties of the cluster i.e. $\delta V/V$ and $\delta M/M$.
We now provide a heuristic argument that provides some intuition as to why the systematic error in $H_0$ leads to the 
largest contribution to  $\sigma_T$.
As a starting point, we write the first term of Eq.~(\ref{TotalError}) as 
$\sum^N_k \sigma^2_k=N\langle\mathcal{JX}\rangle^2\langle\delta \mathcal{J}/\mathcal{J}\rangle^2$.
Here we use the notation $\langle ... \rangle$ to indicate quantities averaged over all the clusters, e.g.
the average integrated exposed J-factor  is
$\langle\mathcal{JX}\rangle = (\mathcal{JX})_T/N$.
 With appropriate substitutions, we can now write
\beq
\sigma_T^2=(\mathcal{JX})_T^2\bigg[\frac{1}{N}\langle\delta J/J\rangle^2+\bigg(2\frac{\delta \text{H}_0}{\text{H}_0}\bigg)^2\bigg].
\eeq

For $N\gg1$, the second term dominates. In our case, we have hundreds to tens of thousands of clusters 
(depending on the selection cuts),  so that this criterion is easily satisfied.
Errors on the Hubble constant are on the few percent level.  
With Eq.~(\ref{galaxyError}), we can estimate  $\langle\delta J/J\rangle$ which depends upon $\delta V/V$ and $\delta M/M$.  
 $\delta M/M$  is on the order of twenty percent for the mass method of the Tully 2MASS Catalog.
 In the case of the velocity,  $\delta V/V$ is on the order of 10 percent\footnote{
 The typical velocity of a galaxy in our study is 5000 km/sec.
 The correct value of $\delta V$ is uncertain, so we  conservatively take it to be 600 km/sec as this high value produces
 the least stringent bounds; we note that $\delta V/V$ is the second most important error after that of the Hubble constant.
 The Local Group moves at 622 km/s relative to the CMB. Frequently the  dispersion is  assumed to be  $\sim$300 km/s, e.g. 
 in cosmological studies using SNe Ia.
There are some old studies of redshift space distortions that find such values, as referenced in a review article by Strauss \& Willick (1995)~\cite{Strauss:1995fz}, p. 325:
``The Fisher {\it et al.} (1994b) analysis also measures the distortions on nonlinear scales to derive the pair-wise velocity dispersion at 100 km/s, $\sigma = 317(+40-49)$ km/s. This is to be compared with the Davis and Peebles (1983b) value of $340(\pm40)$ km/s from the CfA survey, also measured by looking at redshift space distortions."
But in general, one can derive the dispersion from the velocity/matter power spectra, see e.g. 
Hui \& Greene (2006)~\cite{Hui:2005nm}. Sometimes even less is assumed for the dispersion 
(e.g. 150 km/s in Conley {\it et al.} 2011~\cite{2011ApJS..192....1C} and Betoule {\it et al.} 2014~\cite{2014A&A...568A..22B}).  
As a conservative estimate, we take  $\delta V=600$ km/sec.
 For most of our clusters, $V$ is on average on the order of 5000 km/sec.
}

\subsection{Putting Pieces Together}

Now that we have explained the various elements which contribute to our overall likelihood function. We can now 
turn to constraining the properties of dark matter.
Using a delta-log-likelihood approach we can use Fermi-LAT data 
to constrain $\sigmav$ for a given $\mchi$ by treating 
$\beta$, $\gamma$, and $\mathcal{(JX)}$ as nuisance
parameters, which we profile out.  The delta-log-likelihood $\Delta\like$ is given by
\begin{equation} 
    \label{eqn:deltaloglike}
    \Delta\ln\like(m_\chi, \langle\sigma v\rangle)
        \equiv 
    \ln\like(m_\chi, \langle\sigma v\rangle,\dhatbeta,\dhatgamma,\dhatJX)
               - \ln\like(\mchi,\hatsigmav,\hatbeta,\hatgamma,\hatJX \, ,
\end{equation}
where $\hatsigmav$, $\hatJX$, $\hatbeta$, \& $\hatgamma$ are the values of $\sigmav$,
$\mathcal{(JX)}$, $\beta$, \& $\gamma$ that jointly maximize the likelihood at a given $\mchi$.
$\dhatJX$, $\dhatbeta$, and $\dhatgamma$ are the value of $\mathcal{(JX)}$, $\beta$, and $\gamma$ 
that jointly maximize the likelihood for a given $\mchi$ and $\sigmav$.
We should note that the maximum likelihood values for $\beta$ and $\gamma$
are consistent with what we would expect from galactic foreground model of the $\gamma$-rays
from cosmic-rays \cite{Ackermann:2014usa,Acero:2016qlg}.

The 1D confidence intervals in $\sigmav$ at the $n\sigma$ confidence level
are determined by identifying the range of $\sigmav$ such that

\begin{equation} \label{eqn:nsigma}
    \Delta\ln\like(m_\chi, \langle\sigma v\rangle) \le n^2/2 \, .
\end{equation}

In the next section we will present the upper limit of the 2$\sigma$ confidence
intervals (95.4\%~confidence level) for our 3 different selection cuts in the number of clusters. In our analysis, we did not find 
any significant deviations from background.  We found a few regions with
 2 $\sigma$ fluctuations (i.e. consistent with
 purely background). 
 
\begin{figure*}[t!]
    \includegraphics[width=\textwidth, keepaspectratio]{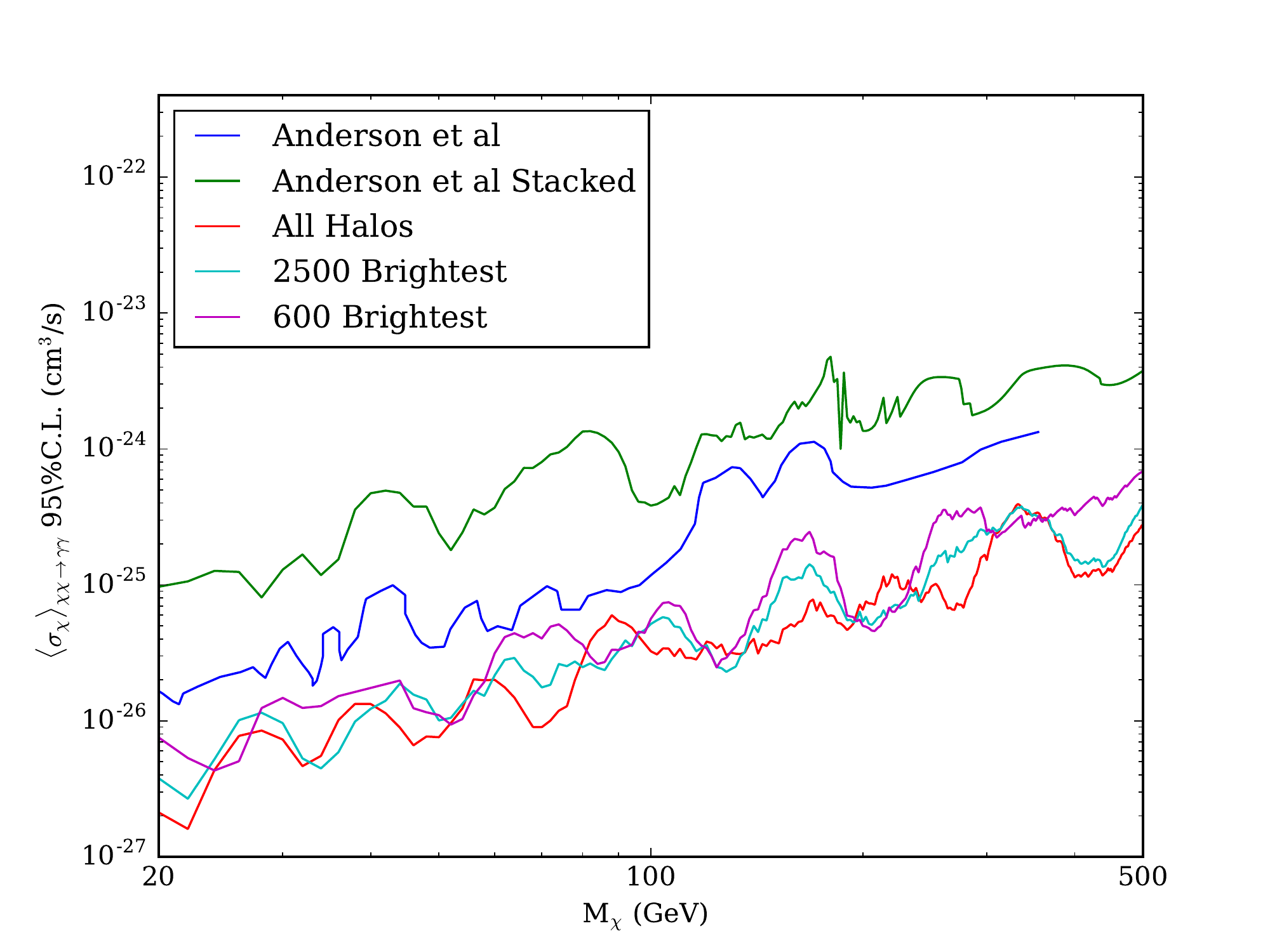}
    \caption{   
        The 95\% confidence upper limits of dark matter cross section as a function of particle mass for $\chi\chi \rightarrow \gamma\gamma$. 
        The figure includes our likelihood analysis for our 3 cases: 
        on all the clusters, the 2500 brightest clusters, and the 600 brightest clusters. 
        In addition, the figure also shows the Anderson {\it et al.} upper confidence bounds \cite{Anderson:2015dpc}.
        Finally, the figure  also includes upper confidence limits on the same clusters as Anderson {\it et al.} using our stacking method. 
        }
           \label{fig:ConfidenceCurve}
\end{figure*}

\section{Results}

 Figure 10 illustrates out main results.  We plot 95\% confidence upper limits on dark matter cross section as a function of particle mass for $\chi\chi \rightarrow \gamma\gamma$. The lowest three curves illustrate the results of our likelihood analysis for our 3 cases: 
        on all the clusters, the 2500 brightest clusters, and the 600 brightest clusters. 
        For comparison, we also show the Anderson {\it et al.} upper limits \cite{Anderson:2015dpc} (prior to our
        work, the strongest bounds from clusters).
        Finally, the figure  also includes upper confidence limits on the same clusters as Anderson {\it et al.} using our stacking method. 
Our results are stronger than previous cluster based results,
using \cite{Anderson:2015dpc} as a recent comparison point, 
but not as strong as previous galactic center based dark matter constraint results \GCDMCITE. 
As expected from Eq.(\ref{eqn:dphidE}), the bounds on the cross section get weaker for higher WIMP mass.

We note the two most significant upward fluctuations are at around 90 GeV and a bump at 350 GeV,
both with a significance of roughly 2 $\sigma$. The fluctuation is slightly more pronounced
in the case where all the clusters are included,  compared to the sub-sample constraints from the 2500 and 600 cluster sets.
Just as a matter of total speculation, we remind the reader that 350 GeV is around half the energy of
the anomalous  bump found at LHC.

A few notes on the relative merits of the different cuts on the clusters.
For most values of DM mass, the constraints are stronger when including more galaxy clusters.
However, at some masses, the constraints are better for smaller subsets of galaxy clusters.
We attribute this to statistical fluctuation. For instance at  $m_\chi \sim 200$ GeV, 
 the 600 halo case has stronger constraints than the all halo case. For the all halo case at this energy,
there are $\sim$70 photons (which we assume are background).  Given this number, for the 600 halo case,
we would expect around 12 photons, but there are only 4 photons. For this case we find that the signal-to-noise ratio is 1.4 times
larger for the 600 halo case and thus the bound is stronger.

The top two curves of Figure 10  contrast 
our stack method with the prior analysis of Anderson etal \cite{Anderson:2015dpc}, which 
 studied only 16 clusters.
 The top two curves in the figure illustrate bounds on the same 16 clusters using our technique (green curve) 
 and the Anderson etal technique (blue curve).  Their bounds are expected to be stronger because they obtain a separate
likelihood for each of the clusters individually whereas we stack the clusters.  In particular, an individual likelihood for each cluster
can incorporate spatial information to reduce the background from point sources and other backgrounds,
whereas in our stacking method we must instead throw out clusters coincident with the 150 brightest Fermi-LAT $\gamma$-ray point sources.
Most of the point source photon energies are below 100 GeV,
and one can see that the biggest difference between the two approaches (green and blue curves) is at these lower energies.
 Indeed our results are weaker by up to an 
order of magnitude, yet do roughly follow the same shape as a function of WIMP mass.
In principle the method of  \cite{Anderson:2015dpc} does provide stronger bounds, and in the future we plan to 
use that approach in studying a few hundred clusters from the Tully catalog.

Our final results (the bottom three curves in the figure) provide far stronger bounds than the previous work of
 \cite{Anderson:2015dpc} due to the simple fact that we are studying 26,000 clusters (rather than the 16 of \cite{Anderson:2015dpc}).

\section{Summary}

In this paper we used $\gamma$-ray observations of some 26,000 galaxy clusters in Pass-7 Fermi Large Area Telescope (LAT) data to place some of the strongest bounds to date on the cross section of dark matter particles for annihilation to a gamma-ray line,
$\chi\chi \rightarrow \gamma\gamma$.  The clusters were selected from the Tully 2MASS Groups catalog.   For each cluster, we defined a region of interest containing 95\% of the expected DM annihilation luminosity.  
We slid a bin of energy range $\sim$ twice the energy resolution of Fermi-LAT 
across our full spectrum of interest.
We then added up all the observed photons for all the clusters for each energy window, 
and compared our observed photon count to the expected count.  
We searched for a bump in the observed photon count above an expected power law background, 
i.e.  a  line (or internal bremsstrahlung \cite{Bergstrom:1989jr,Bringmann:2012vr}) signal at an energy equal to the WIMP mass due to DM annihilation.  
Since no excess above background was found, we used the null signal to place bounds on the DM annihilation cross 
section as a function of WIMP mass for $m_\chi$ between 20 and 500 GeV.
We found $2 \sigma$ upward fluctuations most prominently at $\sim$350 GeV, 
but these are clearly not significant enough to claim detection. 

 We have improved on the previous best limits provided from galaxy clusters by a factor of 5 to 10
(depending on the DM mass).
    Our cluster based constraints are not yet as strong as bounds placed using the Galactic Center, 
    although a less conservative ``boost factor'' from cluster substructure than the one we have chosen
    could strengthen our bounds considerably.
    Our analysis, given this choice of possible boost, is not yet sensitive enough to fully rule out typical realistic DM candidates, especially if the gamma-ray line is not a dominant annihilation mode.

In the analysis in this paper we stacked the clusters.  In the future, a stronger bound may be obtained by performing a different analysis,
namely determining an individual likelihood function for each of several hundred clusters (this approach was previously 
used by  \cite{Anderson:2015dpc} for 16 clusters).  The latter technique
can incorporate spatial information to reduce the background.

Additional work for the future would be to
 repeat our analysis with Fermi-LAT Pass 8 data instead of Pass 7 data, which may yield stronger constraints (the Pass 8 data
 were not available when we started this work).
One could also perform a joint likelihood analysis of both the nearby clusters in conjunction with the Galactic Center. 

Our improvement on the previous limits provided from galaxy clusters by a factor of 5 to 10 shows 
promising future use of galaxy clusters as a viable source for placing further bounds on the particle physics of dark matter.
On general merits,  by looking at clusters of galaxies which have been characterized comprehensively 
by Tully, we now have the strongest constraints on dark matter annihilation to a line in clusters of galaxies. 
In future work, we also plan a similar study on other DM annihilation channels, which will produce a broader range
of photon energies.

\section*{Acknowledgments}
We are grateful for financial support from the Swedish Research
Council (VR) through grant number 2012-2250, and through the Oskar Klein Centre.
We gratefully thank David Spergel, Dragan Huterer, Dan Hooper, and Stephan Zimmer for discussion.

\end{document}